\documentclass[12 pt,twoside,aps,prd,amsmath,amssymb,
tightenlines,showpacs,showkeys,eqsecnum]{revtex4-1}

\usepackage{graphicx}
\usepackage{mathptmx}
\usepackage{bm}
\newcommand {\ve}{\varepsilon}
\newcommand {\pr}{\partial}

\newcommand {\cD}{\cal D}
\newcommand {\bg}{\bar \gamma}
\newcommand {\G}{\Gamma}
\newcommand {\bp}{\bar \psi}
\newcommand {\p}{\psi}

\def\myfigure #1#2#3#4
{\begin{figure}[ht]\begin{center}
\includegraphics[width=#2 \textwidth]{#1.eps}
\parbox[t]{#4\textwidth}{\caption{#3}\label{#1}}
\end{center}\end{figure}}

\def \myfigures #1#2#3#4#5#6#7#8
{\begin{figure}[ht]
    \begin{center}
        \includegraphics[width=#2 \textwidth]{#1.eps}
        \hfill
        \includegraphics[width=#6 \textwidth]{#5.eps}
        \parbox[t]{#4\textwidth}{\caption {#3}\label{#1}}
        \hfill
        \parbox[t]{#8\textwidth}{\caption {#7}\label{#5}}
    \end{center}
\end{figure} }

\begin{document}
\baselineskip -24pt
\title{Nonlinear Spinor Fields in Bianchi type-I spacetime: Problems and Possibilities}
\author{Bijan Saha}
\affiliation{Laboratory of Information Technologies\\
Joint Institute for Nuclear Research\\
141980 Dubna, Moscow region, Russia} \email{bijan@jinr.ru}
\homepage{http://bijansaha.narod.ru}

\begin{abstract}

Within the scope of Bianchi type-I cosmological model we study the
role of spinor field in the evolution of the Universe. It is found
that due to the spinor affine connections the energy momentum tensor
of the spinor becomes non-diagonal, whereas the Einstein tensor is
diagonal. This non-triviality of non-diagonal components of the
energy-momentum tensor imposes some severe restrictions either on
the spinor field or on the metric functions or on both of them. In
case if the restrictions are imposed on the components of spinor
field only, we come to a situation when spinor field becomes
massless and invariants constructed from bilinear spinor forms also
become trivial. Imposing restriction wholly on metric functions we
obtain FRW model, while if the restrictions are imposed both on
metric functions and spinor field components, we come to LRS BI
model. In both cases the system is solved completely. It was found
that if the relation between the pressure and energy density obeys a
barotropic equation of state, only a non-trivial spinor mass can
give rise to a dynamic EoS parameter.
\end{abstract}

\keywords{Spinor field, dark energy, anisotropic cosmological
models, isotropization}

\pacs{98.80.Cq}

\maketitle

\bigskip

\section{Introduction}

The journey of Einstein's General Theory of Relativity in
cosmological area was never a smooth one. The introduction of the
cosmological constant and its' further omission opened the thorny
road from the very beginning. But with many fundamental questions
remaining unanswered and further development and new findings of
observational cosmology lead to the conclusion that Einstein's
General Relativity is not the final theory of gravitational
interactions. These issues come from cosmology and quantum field
theory.  The presence of Big Bang singularity, flatness and
horizontal problems \cite{guth} lead to the fact that the standard
cosmological model \cite{wein} based on GR and the standard model of
particle physics are inadequate to describe the Universe at extreme
regime. The absence of the genuine quantum gravity theory leads to
develop alternative theory of gravity, where, at least, in
semi-classical limits, GR and its positive results could be
recovered.

A fruitful approach in this search is the extended theories of
Gravity (ETG) which have become a sort of paradigm in the study of
gravitational interactions \cite{capoz}. These theories are
essentially based on the corrections and enlargements of Einstein's
theory of Gravity. The paradigm consists of adding higher order
curvature invariants and non-minimally coupled scalar fields into
dynamics resulting from effective action of quantum gravity. An
excellent review on extended theories of gravity can be found in
\cite{capoz1}.

Though the inflationary model \cite{guth,ratra,olive}, described by
a scalar field, known as inflaton, solves the problem of flatness,
isotropy of microwave background radiation and unwanted relics, the
question of where the scalar field comes from and why it undergoes
such a peculiar phase transition from false to right vacuum still
remains unanswered. Moreover, recent observations showed an
accelerated mode of expansion of the present day Universe
\cite{riess,perlmutter}. This leads cosmologists to reconsider
alternative possibilities.

As one of the way out many specialists considered spinor field as an
alternative source. Being related to almost all stable elementary
particles such as proton, electron and neutrino, spinor field,
especially Dirac spin-$1/2$ play a principal role at the microlevel.
However, in cosmology, the role of spinor field was generally
considered to be restricted. Only recently, after some remarkable
works  by different authors
\cite{henneaux,ochs,saha1997a,saha1997b,saha2001a,saha2004a,
saha2004b,saha2006c,saha2006e,saha2007,saha2006d,greene,ribas,souza,kremer},
showing the important role that spinor fields play on the evolution
of the Universe, the situation began to change. This change of
attitude is directly related to some fundamental questions of modern
cosmology:

{\bf (i) Problem of initial singularity:} One of the problems of
modern cosmology is the presence of initial singularity, which means
the finiteness of time. The main purpose of introducing a nonlinear
term in the spinor field Lagrangian is to study the possibility of
the elimination of initial singularity. In a number of papers
\cite{saha1997a,saha1997b,saha2001a,saha2004a,saha2004b} it was
shown that the introduction of spinor field with a suitable
nonlinearity into the system indeed gives rise to singularity-free
models of the Universe. It should be noted that the singularity-free
solutions in the papers mentioned were obtained at the expense of
dominant energy condition. Problem of singularity and its possible
elimination exploiting spinor field were discussed in
\cite{PopPLB,PopPRD,PopGREG,FabIJTP}.

{\bf  (ii) problem of isotropization:} Although the Universe seems
homogenous and isotropic at present, it does not necessarily mean
that it is also suitable for a description of the early stages of
the development  of the Universe and there are no observational data
guaranteeing the isotropy in the era prior to the recombination. In
fact, there are theoretical arguments that support the existence of
an anisotropic phase that approaches an isotropic one \cite{misner}.
The observations from Cosmic Background Explorer's differential
radiometer have detected and measured cosmic microwave background
anisotropies in different angular scales. Recently Planck has
compiled the most detailed map of the cosmic microwave background
ever created. The new map renews our understanding of the Universe's
composition and evolution. The image of the cosmic microwave
background (CMB) composed from the lights, imprinted on the sky when
the Universe was just 380 000 years old shows tiny temperature
fluctuations that correspond to regions of slightly different
densities at very early times. These anisotropies are supposed to
hide in their fold the entire history of cosmic evolution dating
back to the recombination era and are being considered as indicative
of the geometry and the content of the universe. There is widespread
consensus among the cosmologists that cosmic microwave background
anisotropies in small angular scales have the seeds of all future
structure: the stars and galaxies of today. It was found that the
introduction of nonlinear spinor field accelerates the
isotropization process of the initially anisotropic Universe
\cite{saha2001a,saha2004a,saha2006c,PopPRD}.

{\bf  (iii) late time acceleration of the Universe:} Some recent
experiments detected an accelerated mode of expansion of the
Universe \cite{riess,perlmutter}. Detection and further experimental
reconfirmation of current cosmic acceleration pose to cosmology a
fundamental task of identifying and revealing the cause of such
phenomenon. This fact can be reconciled with the theory if one
assumes that the Universe id mostly filled with so-called dark
energy. This form of matter (energy) is not observable in laboratory
and it does not interact with electromagnetic radiation. These facts
played decisive role in naming this object. In contrast to dark
matter, dark energy is uniformly distributed over the space, does
not intertwine under the influence of gravity in all scales and it
has a strong negative pressure of the order of energy density. Based
on these properties, cosmologists have suggested a number of dark
energy models those are able to explain the current accelerated
phase of expansion of the Universe. In this connection a series of
papers appeared recently in the literature, where a spinor field was
considered as an alternative model for dark energy \cite{ribas,
saha2006d,saha2006e,saha2007,PopGREG}.

It should be noted that most of the works mentioned above were
carried out within the scope of Bianchi type-I cosmological model.
Results obtained using a spinor field as a source of Bianchi type-I
cosmological field can be summed up as follows: A suitable choice of
spinor field nonlinearity\\

(i) {\it accelerates the isotropization process}
\cite{saha2001a,saha2004a,saha2006c};\\

(ii) {\it gives rise to a singularity-free Universe}
\cite{saha2001a,saha2004a,saha2004b,saha2006c};\\

(iii) {\it generates late time acceleration}
\cite{ribas,saha2006e,saha2007,saha2006d,souza}.

Given the role that spinor field can play in the evolution of the
Universe, question that naturally pops up is, if the spinor field
can redraw the picture of evolution caused by perfect fluid and dark
energy, is it possible to simulate perfect fluid and dark energy by
means of a spinor field? Affirmative answer to this question was
given in the a number of papers
\cite{krechet,saha2010a,saha2010b,saha2011,saha2012}. In those
papers spinor description of matter such as perfect fluid and dark
energy was given and the evolution of the Universe given by
different Bianchi models was thoroughly studied. In almost all the
papers the spinor field was considered to be time-dependent
functions and its energy-momentum tensor was given by the diagonal
elements only. Some latest study shows that due to the specific
connection with gravitational field the energy-momentum tensor of
the spinor field possesses non-trivial non-diagonal components as
well. In this paper we study the role of non-diagonal components of
the energy-momentum tensor of the spinor field in the evolution of
the Universe. To our knowledge such study was never done previously.
In section II we give the spinor field Lagrangian in details. In
section III the system of Einstein-Dirac equations is solved for BI
metric without engaging the non-diagonal components of
energy-momentum tensor as it was done in previous works of many
authors. In section IV we analyze the role of non-diagonal
components of energy-momentum tensor on the evolution of the
Universe.

It should also be noted that recently inflation has been studied
within the scope of spinor theory as well. Basic theory of dark
spinor inflation is presented in \cite{CB,Mota}. The ELKO field in
interaction through contortion with its own spin density was studied
in \cite{FabGRG}, that was further developed in \cite{ELKO}.
Conformal coupling of dark spinor field to gravity within the scope
of FRW model was studied in \cite{Lee}.

Recently many authors studied the Dirac spinors within the scope of
different cosmological models with
torsion\cite{PopPLB,FabIJTP,ELKO,FabJMP,PopPRD,PopGREG}. Dirac
spinors in Bianchi type-I $f(R)$ cosmology with torsion was studied
in \cite{FabJMP}, where it was shown that the dynamic behavior of
the universe depends on the particular choice of the function
$f(R)$. In this paper tt was highlighted that, despite the
anisotropic background the Einstein tensor is diagonal, whereas,
because of intrinsic feature of spinor field, the energy tensor is
non-diagonal. Dirac field equations coupled to electrodynamics and
torsion fields were investigated in \cite{FabIJTP}. It was shown
that minimal coupling between the torsion tensor and Dirac spinors
generates  a spin-spin interaction \cite{PopPRD}. In \cite{PopGREG}
it was shown that the spacetime torsion, generated by Dirac spinors,
induces gravitational repulsion. Nonsingular Dirac particles in
spacetime with torsion were studied in \cite{PopPLB}.

\section{Basic equation}

Let us consider the case when the anisotropic space-time is filled
with nonlinear spinor field. The corresponding action can be given
by
\begin{equation}
{\cal S}(g; \psi, \bp) = \int\, L \sqrt{-g} d\Omega \label{action}
\end{equation}
with
\begin{equation}
L= L_{\rm g} + L_{\rm sp}. \label{lag}
\end{equation}
Here $L_{\rm g}$ corresponds to the gravitational field
\begin{equation}
L_{\rm g} = \frac{R}{2\kappa}, \label{lgrav}
\end{equation}
where $R$ is the scalar curvature, $\kappa = 8 \pi G$, with G being
Einstein's gravitational constant and $L_{\rm sp}$ is the spinor
field Lagrangian.

\subsection{Gravitational field}

The gravitational field in our case is given by a Bianchi type-I
anisotropic space time:
\begin{equation}
ds^2 = dt^2 - a_1^{2} \,dx^{2} - a_2^{2} \,dy^{2} - a_3^{2}\,dz^2,
\label{bi}
\end{equation}
with $a_1,\,a_2$ and $a_3$ being the functions of time only. It is
the simplest anisotropic model of space-time. The reason for
considering anisotropic model lays on the fact that though an
isotropic FRW model describes the present day Universe with great
accuracy, there are both theoretical arguments and observational
data suggesting the existence of an anisotropic phase in the remote
past.

The nonzero components of the Einstein tensor corresponding to the
metric \eqref{bi} are
\begin{subequations}
\label{ET}
\begin{eqnarray}
G_1^1 &=& - \frac{\ddot a_2}{a_2} - \frac{\ddot a_3}{a_3} -
\frac{\dot a_2}{a_2}\frac{\dot a_3}{a_3}, \label{ET11}\\
G_2^2 &=& - \frac{\ddot a_3}{a_3} - \frac{\ddot a_1}{a_1} -
\frac{\dot a_3}{a_3}\frac{\dot a_1}{a_1}, \label{ET22}\\
G_3^3 &=& - \frac{\ddot a_1}{a_1} - \frac{\ddot a_2}{a_2} -
\frac{\dot a_1}{a_1}\frac{\dot a_2}{a_2}, \label{ET33}\\
G_1^1 &=& - \frac{\dot a_1}{a_1}\frac{\dot a_2}{a_2} - \frac{\dot
a_2}{a_2}\frac{\dot a_3}{a_3}- \frac{\dot a_3}{a_3}\frac{\dot
a_1}{a_1}. \label{ET00}
\end{eqnarray}
\end{subequations}

\subsection{Spinor field}

For a spinor field $\p$, the symmetry between $\p$ and $\bp$ appears
to demand that one should choose the symmetrized Lagrangian
\cite{kibble}. Keeping this in mind we choose the spinor field
Lagrangian as \cite{saha2001a}:
\begin{equation}
L_{\rm sp} = \frac{\imath}{2} \biggl[\bp \gamma^{\mu} \nabla_{\mu}
\psi- \nabla_{\mu} \bar \psi \gamma^{\mu} \psi \biggr] - m_{\rm sp}
\bp \psi - F, \label{lspin}
\end{equation}
where the nonlinear term $F$ describes the self-interaction of a
spinor field and can be presented as some arbitrary functions of
invariants generated from the real bilinear forms of a spinor field.
Here for simplicity we consider the case when $F = F(S)$ with $S=
\bp \psi$. Here $\nabla_\mu$ is the covariant derivative of spinor
field:
\begin{equation}
\nabla_\mu \psi = \frac{\partial \psi}{\partial x^\mu} -\G_\mu \psi,
\quad \nabla_\mu \bp = \frac{\partial \bp}{\partial x^\mu} + \bp
\G_\mu, \label{covder}
\end{equation}
with $\G_\mu$ being the spinor affine connection. In \eqref{lspin}
$\gamma$'s are the Dirac matrices in curve space-time and obey the
following algebra
\begin{equation}
\gamma^\mu \gamma^\nu + \gamma^\nu \gamma^\mu = 2 g^{\mu\nu}
\label{al}
\end{equation}
and are connected with the flat space-time Dirac matrices $\bg$ in
the following way
\begin{equation}
 g_{\mu \nu} (x)= e_{\mu}^{a}(x) e_{\nu}^{b}(x) \eta_{ab},
\quad \gamma_\mu(x)= e_{\mu}^{a}(x) \bg_a, \label{dg}
\end{equation}
where $\eta_{ab}= {\rm diag}(1,-1,-1,-1)$ and $e_{\mu}^{a}$ is a set
of tetrad 4-vectors. The spinor affine connection matrices $\G_\mu
(x)$ are uniquely determined up to an additive multiple of the unit
matrix by the equation
\begin{equation}
\frac{\pr \gamma_\nu}{\pr x^\mu} - \G_{\nu\mu}^{\rho}\gamma_\rho -
\G_\mu \gamma_\nu + \gamma_\nu \G_\mu = 0, \label{afsp}
\end{equation}
with the solution
\begin{equation}
\Gamma_\mu = \frac{1}{4} \bg_{a} \gamma^\nu \partial_\mu e^{(a)}_\nu
- \frac{1}{4} \gamma_\rho \gamma^\nu \Gamma^{\rho}_{\mu\nu},
\label{sfc}
\end{equation}

\subsection{Field equations}

Variation of \eqref{action} with respect to the metric function
$g_{\mu \nu}$ gives the Einstein field equation
\begin{equation}
G_\mu^\nu = R_\mu^\nu - \frac{1}{2} \delta_\mu^\nu R = -\kappa
T_\mu^\nu, \label{EEg}
\end{equation}
where $R_\mu^\nu$ and $R$ are the Ricci tensor and Ricci scalar,
respectively. Here $T_\mu^\nu$ is the energy momentum tensor of the
spinor field.

Varying \eqref{lspin} with respect to $\bp (\psi)$ one finds the
spinor field equations:
\begin{subequations}
\label{speq}
\begin{eqnarray}
\imath\gamma^\mu \nabla_\mu \psi - m_{\rm sp} \psi -  F_S  \psi &=&0, \label{speq1} \\
\imath \nabla_\mu \bp \gamma^\mu +  m_{\rm sp} \bp + 2 F_S \bp &=&
0. \label{speq2}
\end{eqnarray}
\end{subequations}
Here we denote $F_S = dF/dS$.

\subsection{Energy momentum tensor of the spinor field}

The energy-momentum tensor of the spinor field is given by
\begin{equation}
T_{\mu}^{\rho}=\frac{\imath}{4} g^{\rho\nu} \biggl(\bp \gamma_\mu
\nabla_\nu \psi + \bp \gamma_\nu \nabla_\mu \psi - \nabla_\mu \bar
\psi \gamma_\nu \psi - \nabla_\nu \bp \gamma_\mu \psi \biggr) \,-
\delta_{\mu}^{\rho} L_{\rm sp} \label{temsp}
\end{equation}
where $L_{\rm sp}$ in view of \eqref{speq} can be rewritten as
\begin{eqnarray}
L_{\rm sp} & = & \frac{\imath}{2} \bigl[\bp \gamma^{\mu}
\nabla_{\mu} \psi- \nabla_{\mu} \bar \psi \gamma^{\mu} \psi \bigr] -
m_{\rm sp} \bp \psi - F(K)
\nonumber \\
& = & \frac{\imath}{2} \bp [\gamma^{\mu} \nabla_{\mu} \psi - m_{\rm
sp} \psi] - \frac{\imath}{2}[\nabla_{\mu} \bar \psi \gamma^{\mu} +
m_{\rm sp} \bp] \psi
- F(S),\nonumber \\
& = &  S F_S - F(S). \label{lspin01}
\end{eqnarray}

Then in view of \eqref{covder} the energy-momentum tensor of the
spinor field can be written as
\begin{eqnarray}
T_{\mu}^{\,\,\,\rho}&=&\frac{\imath}{4} g^{\rho\nu} \bigl(\bp
\gamma_\mu
\partial_\nu \psi + \bp \gamma_\nu \partial_\mu \psi -
\partial_\mu \bar \psi \gamma_\nu \psi - \partial_\nu \bp \gamma_\mu
\psi \bigr)\nonumber\\
& - &\frac{\imath}{4} g^{\rho\nu} \bp \bigl(\gamma_\mu \G_\nu +
\G_\nu \gamma_\mu + \gamma_\nu \G_\mu + \G_\mu \gamma_\nu\bigr)\psi
 \,- \delta_{\mu}^{\rho} \bigl( S F_S - F(S)\bigr). \label{temsp0}
\end{eqnarray}
As is seen from \eqref{temsp0}, is case if for a given metric
$\G_\mu$'s are different, there arise nontrivial non-diagonal
components of the energy momentum tensor.

From the \eqref{sfc} one finds the following expressions for spinor
affine connections:
\begin{equation}
\G_0 = 0, \quad \G_1 = \frac{\dot a_1}{2} \bg^1 \bg^0, \quad \G_2 =
\frac{\dot a_2}{2} \bg^2 \bg^0, \quad \G_3 = \frac{\dot a_3}{2}
\bg^3 \bg^0. \label{sac123}
\end{equation}

We consider the case when the spinor field depends on $t$ only. Then
from \eqref{temsp0} one finds
\begin{subequations}
\label{Ttotext}
\begin{eqnarray}
T_0^0 & = & \frac{\imath}{2} g^{00} \bigl(\bp \gamma_0
\dot \psi - \dot \bp \gamma_0 \psi\bigr) - L_{\rm sp},\label{Ttot00ext}\\
T_1^1 & = & -\frac{\imath}{2} g^{11} \bp \bigl(\gamma_1 \G_1 + \G_1 \gamma_1) \psi - L_{\rm sp},\label{Ttot11ext}\\
T_2^2 & = & -\frac{\imath}{2} g^{22} \bp \bigl(\gamma_2 \G_2 + \G_2 \gamma_2) \psi- L_{\rm sp},\label{Ttot22ext}\\
T_3^3 & = & -\frac{\imath}{2} g^{33} \bp \bigl(\gamma_3 \G_3 + \G_3 \gamma_3) \psi- L_{\rm sp},\label{Ttot33ext}\\
T_1^0 & = & \frac{\imath}{4} g^{00} \bigl(\bp \gamma_1 \dot \psi -
\dot \bp \gamma_1 \psi\bigr) - \frac{\imath}{4} g^{00} \bp
\bigl(\gamma_0 \G_1 + \G_1 \gamma_0\bigr) \psi, \label{Ttot01ext}\\
T_2^0 & = & \frac{\imath}{4} g^{00} \bigl(\bp \gamma_2 \dot \psi -
\dot \bp \gamma_2 \psi\bigr) - \frac{\imath}{4} g^{00} \bp
\bigl(\gamma_0 \G_2 + \G_2 \gamma_0\bigr) \psi, \label{Ttot02ext}\\
T_3^0 & = & \frac{\imath}{4} g^{00} \bigl(\bp \gamma_3 \dot \psi -
\dot \bp \gamma_3 \psi\bigr) - \frac{\imath}{4} g^{00} \bp
\bigl(\gamma_0 \G_3 + \G_3 \gamma_0\bigr) \psi, \label{Ttot03ext}\\
T_2^1 & = & -\frac{\imath}{4} g^{11} \bp \bigl(\gamma_2 \G_1 + \G_1
\gamma_2 + \gamma_1 \G_2 + \G_2 \gamma_1\bigr) \psi, \label{Ttot12ext}\\
T_3^2 & = & -\frac{\imath}{4} g^{22} \bp \bigl(\gamma_3 \G_2 + \G_2
\gamma_3 + \gamma_2 \G_3 + \G_3 \gamma_2\bigr) \psi, \label{Ttot23ext}\\
T_3^1 & = &  -\frac{\imath}{4} g^{11} \bp \bigl(\gamma_3 \G_1 + \G_1
\gamma_3 + \gamma_1 \G_3 + \G_3 \gamma_1\bigr) \psi.
\label{Ttot31ext}
\end{eqnarray}
\end{subequations}

In this case after a little manipulations from \eqref{temsp0} for
the nontrivial components of the energy momentum tensor one finds
\cite{BIreex}:

\begin{subequations}
\label{Ttot}
\begin{eqnarray}
T_0^0 & = &  m_{\rm sp} S + F(S),\label{Ttot00}\\
T_1^1 & = & T_2^2 = T_3^3 = F(S) - S F_S,\label{Ttot11}\\
T_2^1 & = & -\frac{\imath}{4} \frac{a_2}{a_1} \biggl(\frac{\dot
a_1}{a_1} - \frac{\dot a_2}{a_2}\biggr)  \bp \bg^1 \bg^2 \bg^0 \psi, \label{Ttot12}\\
T_3^2 & = & -\frac{\imath}{4} \frac{a_3}{a_2} \biggl(\frac{\dot
a_2}{a_2} - \frac{\dot a_3}{a_3}\biggr)  \bp \bg^2 \bg^3 \bg^0 \psi, \label{Ttot23}\\
T_3^1 & = &  -\frac{\imath}{4} \frac{a_3}{a_1} \biggl(\frac{\dot
a_3}{a_3} - \frac{\dot a_1}{a_1}\biggr)  \bp \bg^3 \bg^1 \bg^0 \psi.
\label{Ttot31}
\end{eqnarray}
\end{subequations}

As one sees from \eqref{Ttotext} and \eqref{Ttot} the non-triviality
of non-diagonal components of the energy momentum tensors, namely
$T_2^1$, $T_3^2$ and $T_3^1$ is directly connected with the affine
spinor connections $\G_i$'s.

\section{Solution to the field equations}

In this section we solve the field equations. Let us begin with the
spinor field equations. In view of \eqref{covder} and \eqref{sac123}
the spinor field equation \eqref{speq1} takes the form
\begin{subequations}
\label{SF1}
\begin{eqnarray}
\imath \gamma^0 \bigl(\dot \psi + \frac{1}{2}\frac{\dot V}{V}
\psi\bigr) - m_{\rm sp} \psi - S F_S  \psi &=& 0. \label{speq1p}\\
\imath \gamma^0 \bigl(\dot \psi + \frac{1}{2}\frac{\dot V}{V}
\psi\bigr) + m_{\rm sp} \psi + S F_S  \bp &=& 0, \label{speq2p}
\end{eqnarray}
\end{subequations}
where we define the volume scale as
\begin{equation}
V = a_1 a_2 a_3. \label{VDef}
\end{equation}
From \eqref{SF1} one easily finds
\begin{equation}
\dot S + \frac{\dot V}{V} S = 0, \label{VS}
\end{equation}
with the solution
\begin{equation}
S = \frac{V_0}{V}, \quad V_0 = {\rm const}. \label{VS0}
\end{equation}

As we have already mentioned, $\psi$ is a function of $t$ only. We
consider the 4-component spinor field given by
\begin{eqnarray}
\psi = \left(\begin{array}{c} \psi_1\\ \psi_2\\ \psi_3 \\
\psi_4\end{array}\right). \label{psi}
\end{eqnarray}
Taking into account that $\phi_i =\sqrt{V} \psi_i$ and defining
${\cD} =  S F_S$ and inserting \eqref{psi} into \eqref{speq1p} in
this case we find
\begin{subequations}
\label{speq1pfg}
\begin{eqnarray}
\dot \phi_1 + \imath {\cD} \phi_1 &=& 0, \label{ph1}\\
\dot \phi_2 + \imath {\cD} \phi_2 &=& 0, \label{ph2}\\
\dot \phi_3 - \imath {\cD} \phi_3 &=& 0, \label{ph3}\\
\dot \phi_4 - \imath {\cD} \phi_4 &=& 0, \label{ph4}.
\end{eqnarray}
\end{subequations}
Here we also consider the massless spinor field setting $m_{\rm sp}
= 0.$ The foregoing system of equations can be easily solved.
Finally for the spinor field we obtain
\begin{subequations}
\label{psinl}
\begin{eqnarray}
\psi_1(t) &=& (C_1/\sqrt{V}) \exp{\Biggl(-i\int {\cD} dt\Biggr)},\\
\psi_2(t) &=& (C_2/\sqrt{V}) \exp{\,\Biggl(-i\int {\cD} dt\Biggr)},\\
\psi_3(t) &=& (C_3/\sqrt{V}) \exp{\,\Biggl(i\int  {\cD} dt\Biggr)},\\
\psi_4(t) &=& (C_4/\sqrt{V}) \exp{\,\Biggl(i\int  {\cD} dt\Biggr)},
\end{eqnarray}
\end{subequations}
with $C_1,\,C_2,\,C_3,\,C_4$ being the integration constants and
related to $V_0$ as
$$C_1^* C_1 + C_2^* C_2 - C_3^* C_3 - C_4^* C_4 = V_0.$$

Thus we see that the components of the spinor field are in some
functional dependence of $V$.

Let us now solve the gravitational field equations. On account of
\eqref{ET} and \eqref{Ttot} the system of Einstein field equations
\eqref{EEg} takes the form
\begin{subequations}
\label{BIE}
\begin{eqnarray}
\frac{\ddot a_2}{a_2} +\frac{\ddot a_3}{a_3} + \frac{\dot
a_2}{a_2}\frac{\dot a_3}{a_3}&=&  \kappa \bigl( F(S) - S F_S\bigr),\label{11bi}\\
\frac{\ddot a_3}{a_3} +\frac{\ddot a_1}{a_1} + \frac{\dot
a_3}{a_3}\frac{\dot a_1}{a_1}&=& \kappa \bigl( F(S) - S F_S\bigr),\label{22bi}\\
\frac{\ddot a_1}{a_1} +\frac{\ddot a_2}{a_2} + \frac{\dot
a_1}{a_1}\frac{\dot a_2}{a_2}&=&  \kappa \bigl( F(S) - S F_S\bigr),\label{33bi}\\
\frac{\dot a_1}{a_1}\frac{\dot a_2}{a_2} +\frac{\dot
a_2}{a_2}\frac{\dot a_3}{a_3} +\frac{\dot a_3}{a_3}\frac{\dot
a_1}{a_1}&=& \kappa \bigl(m_{\rm sp} S + F(S)\bigr), \label{00bi}
\end{eqnarray}
\end{subequations}
together with the additional constrains
\begin{subequations}
\label{AC}
\begin{eqnarray}
\biggl(\frac{\dot
a_1}{a_1} - \frac{\dot a_2}{a_2}\biggr)  \bp \bg^1 \bg^2 \bg^0 \psi &=& 0,\label{12ac}\\
 \biggl(\frac{\dot
a_2}{a_2} - \frac{\dot a_3}{a_3}\biggr)  \bp \bg^2 \bg^3 \bg^0 \psi &=& 0,\label{23ac}\\
\biggl(\frac{\dot a_3}{a_3} - \frac{\dot a_1}{a_1}\biggr)  \bp \bg^3
\bg^1 \bg^0 \psi &=& 0.\label{33ac}
\end{eqnarray}
\end{subequations}

From \eqref{AC} we see, it is possible to consider the case when we
impose restrictions on the spinor field or they will be imposed on
the metric functions. In what follows we will study the both
situations in details. Note that in this case thanks to the Einstein
equations the non-diagonal components of the energy momentum tensor
become zero. But we can also set them zero in order to simulate
different kind of fluid and dark energy those have only non-zero
diagonal components. It should be noted that additional constrains
analogous to \eqref{AC} was also found it \cite{FabJMP}.

\subsection{restrictions on spinor field}

Let us first consider the case when the non-diagonal components of
the energy momentum tensor impose restrictions on the spinor field.
In this case from \eqref{AC} we obtain
\begin{equation}
\bp \bg^1 \bg^2 \bg^0 \psi = \bp \bg^3 \bg^1 \bg^0 \psi = \bp \bg^2
\bg^3 \bg^0 \psi = 0. \label{spinor123}
\end{equation}
The components of the spinor field in this case undergo some
changes. It is found that in this case the integration constants in
\eqref{psinl} should obey
\begin{subequations}
\label{spcons}
\begin{eqnarray}
C_1^* C_1 - C_2^* C_2 - C_3^* C_3 + C_4^* C_4 &=& 0,\label{spcons1}\\
C_1^* C_2 - C_2^* C_1 - C_3^* C_4 + C_4^* C_3 &=& 0,
\label{spcons2}\\
C_1^* C_2 + C_2^* C_1 - C_3^* C_4 - C_4^* C_3 &=& 0, \label{spcons3}\\
C_1^* C_1 + C_2^* C_2 - C_3^* C_3 - C_4^* C_4 &=& V_0,
\label{spcons4}
\end{eqnarray}
\end{subequations}
which gives
\begin{subequations}
\label{spconsnew}
\begin{eqnarray}
C_1^* C_2 - C_3^* C_4 &=&  C_2^* C_1 - C_4^* C_3 = 0,
\label{spconsnew2}\\
C_1^* C_1 - C_3^* C_3 &=&  C_2^* C_2  - C_4^* C_4 = \frac{V_0}{2}.
\label{spconsnew4}
\end{eqnarray}
\end{subequations}

In this case solving the Einstein equation on account of the fact
that $T_1^1 = T_2^2 = T_3^3$ for the metric functions one finds
\cite{saha2001a}
\begin{eqnarray}
a_i = D_i V^{1/3} \exp{\Bigl(X_i \int \frac{dt}{V}\Bigr)}, \quad
\prod_{i=1}^{3} D_i = 1, \quad     \sum_{i=1}^{3} X_i = 0,
\label{metricf}
\end{eqnarray}
with $D_i$ and $X_i$ being the integration constants. Thus we see
that the metric functions can be expressed in terms of $V$.

Summation of \eqref{11bi}, \eqref{22bi}, \eqref{33bi} and 3 times
\eqref{00bi} leads to the equation for $V$  \cite{saha2001a}
\begin{eqnarray}
\ddot V = \frac{3 \kappa}{2} (T_0^0 + T_1^1) V = \frac{3 \kappa}{2}
(m_{\rm sp} S + 2 F(S) - S F_S) V. \label{detvbi}
\end{eqnarray}
Since $S$ is a function of $V$ the right hand side of \eqref{detvbi}
is a function of $V$ as well, hence can be solved in quadrature.
Given the concrete form of view one finds the solution for $V$. Here
we consider the cases, when the spinor field describes different
kinds of well-known fluid and dark energy.

Let us now recall that in the unified nonlinear spinor theory of
Heisenberg, the massive term remains absent, and according to
Heisenberg, the particle mass should be obtained as a result of
quantization of spinor prematter~ \cite{massless,massless1}. In the
nonlinear generalization of classical field equations, the massive
term does not possess the significance that it possesses in the
linear one, as it by no means defines total energy (or mass) of the
nonlinear field system. Moreover, it was established that only a
massless spinor field with the Lagrangian \eqref{lspin} describes
perfect fluid from phantom to ekpyrotic matter
\cite{krechet,saha2010a,saha2010b,saha2011,saha2012}. Thus without
losing the generality we can consider the massless spinor field
putting $m_{\rm sp}\,=\,0.$

Let us consider the case when the spinor field Lagrangian
\eqref{lspin} describes a berotropic fluid. Inserting \eqref{Ttot00}
and \eqref{Ttot11} into the barotropic equation of state
\begin{equation}
p = W \ve, \label{beos}
\end{equation}
where $W$ is a constant, one finds
\begin{equation}
S F_S  = (1 + W) F(S), \label{beos1}
\end{equation}
with the solution
\begin{equation}
F(S) = \lambda S^{(1+W)}, \quad \lambda = {\rm const.} \label{sol1}
\end{equation}
Depending on the value of $W$ \eqref{beos} describes perfect fluid
from phantom to ekpyrotic matter, namely
\begin{subequations}
\label{zeta}
\begin{eqnarray}
W &=& 0, \qquad \qquad \qquad {\rm (dust)},\\
W &=& 1/3, \quad \qquad \qquad{\rm (radiation)},\\
W &\in& (1/3,\,1), \quad \qquad\,\,{\rm (hard\,\,Universe)},\\
W &=& 1, \quad \qquad \quad \qquad {\rm (stiff \,\,matter)},\\
W &\in& (-1/3,\,-1), \quad \,\,\,\,{\rm (quintessence)},\\
W &=& -1, \quad \qquad \quad \quad{\rm (cosmological\,\, constant)},\\
W &<& -1, \quad \qquad \quad \quad{\rm (phantom\,\, matter)},\\
W &>& 1, \quad \qquad \quad \qquad{\rm (ekpyrotic\,\, matter)}.
\end{eqnarray}
\end{subequations}

In account of it the spinor field Lagrangian now reads
\begin{equation}
L = \frac{i}{2} \biggl[\bp \gamma^{\mu} \nabla_{\mu} \psi-
\nabla_{\mu} \bar \psi \gamma^{\mu} \psi \biggr] - \lambda
S^{(1+W)}. \label{lspin1}
\end{equation}
Thus a massless spinor field with the Lagrangian \eqref{lspin1}
describes perfect fluid from phantom to ekpyrotic matter. Here the
constant of integration $\lambda$ can be viewed as constant of
self-coupling. A detailed analysis of this study was given in
\cite{krechet,saha2010a,saha2010b,saha2011}.

In case of \eqref{lspin1} we have
\begin{subequations}
\label{emtquint}
\begin{eqnarray}
T_0^0 &=& \varepsilon = \lambda S^{(1+W)}, \label{emtquint0}\\
T_1^1 &=& - p = - W \varepsilon = -W \lambda S^{(1+W)}.
\label{emtquint1}
\end{eqnarray}
\end{subequations}
Eq. \eqref{detvbi} then takes the form
\begin{equation}
\ddot V = \frac{3\kappa}{2}  \lambda V_0^{1+W} (1-W) V^{-W},
\label{Vquint}
\end{equation}
with the solution in quadrature
\begin{equation}
\int\frac{dV}{\sqrt{3 \kappa \lambda V_0^{1+W} V^{1-W} + C_1}} = t +
t_0. \label{Vquintquad}
\end{equation}
Here $C_1$ and $t_0$ are the integration constants.

Let us consider the case when the spinor field describes a Chaplygin
gas described by a equation of state
\begin{equation}
p = -A/\ve^\alpha, \label{chap}
\end{equation}
where $A$ is a positive constant and $0 \le \alpha \le 1$. Then in
case of a massless spinor field for $F$ one finds
\begin{equation}
\frac{F^\alpha dF}{F^{1+\alpha} - A} = \frac{dS}{S}, \label{eqq}
\end{equation}
with the solution \cite{saha2010a,saha2010b,saha2011}
\begin{equation}
F = \bigl(A + \lambda S^{(1+\alpha)}\bigr)^{1/(1+\alpha)}.
\label{chapsp}
\end{equation}
The Spinor field Lagrangian in this case takes the form
\begin{equation}
L = \frac{i}{2} \biggl[\bp \gamma^{\mu} \nabla_{\mu} \psi-
\nabla_{\mu} \bar \psi \gamma^{\mu} \psi \biggr] - \bigl(A + \lambda
S^{(1+\alpha)}\bigr)^{1/(1+\alpha)}. \label{lspin2}
\end{equation}

In this case we have
\begin{subequations}
\label{emtchap}
\begin{eqnarray}
T_0^0 &=& \varepsilon = \bigl(A + \lambda S^{(1+\alpha)}\bigr)^{1/(1+\alpha)}, \label{emtchap0}\\
T_1^1 &=& - p =  A/\varepsilon^\alpha =  A/\bigl(A + \lambda
S^{(1+\alpha)}\bigr)^{\alpha/(1+\alpha)}. \label{emtchap1}
\end{eqnarray}
\end{subequations}

The equation for $V$ now reads
\begin{equation}
\ddot V = \frac{3\kappa}{2} \Biggl[ \bigl(AV^{1+\alpha} + \lambda
V_0^{1+\alpha}\bigr)^{1/(1+\alpha)} + A
V^{1+\alpha}/\bigl(AV^{1+\alpha} + \lambda
V_0^{1+\alpha}\bigr)^{\alpha/(1+\alpha)}\Biggr], \label{Vchap}
\end{equation}
with the solution
\begin{equation}
\int \frac{dV}{\sqrt{C_1 + 3 \kappa V \bigl(AV^{1+\alpha} + \lambda
V_0^{1+\alpha}\bigr)^{1/(1+\alpha)}}} = t + t_0, \quad C_1 = {\rm
const}. \quad t_0 = {\rm const}. \label{Vchapquad}
\end{equation}
Inserting $\alpha = 1$ we come to the result obtained in
\cite{saha2005}.

We also consider a modified Chaplygin gas given by the equation of
state
\begin{equation}
p = W \ve - A/\ve^\alpha, \label{mchap}
\end{equation}
with $W$ is a constant, $A > 0$ and $0 \le \alpha \le 1$. Inserting
$p = S F_S - F$ and $\ve = F$ (we consider here massless spinor
field) we get
\begin{equation}
\frac{F^\alpha dF}{F^{1+\alpha} - A/(1+W)} = (1+W)\frac{dS}{S},
\label{eqmq}
\end{equation}
with the solution
\begin{equation}
F = \biggl(\frac{A}{1+W} + \lambda
S^{(1+\alpha)(1+W)}\biggr)^{1/(1+\alpha)}. \label{mchapsp}
\end{equation}
The Spinor field Lagrangian in this case takes the form
\begin{equation}
L = \frac{i}{2} \biggl[\bp \gamma^{\mu} \nabla_{\mu} \psi-
\nabla_{\mu} \bar \psi \gamma^{\mu} \psi \biggr] -
\biggl(\frac{A}{1+W} + \lambda
S^{(1+\alpha)(1+W)}\biggr)^{1/(1+\alpha)}. \label{lspin2m}
\end{equation}
In this case we have In this case we have
\begin{subequations}
\label{emtmchap}
\begin{eqnarray}
T_0^0 &=& \varepsilon = \biggl(\frac{A}{1+W} + \lambda S^{(1+\alpha)(1+W)}\biggr)^{1/(1+\alpha)}, \label{emtmchap0}\\
T_1^1 &=& - p = - W \varepsilon + A/\varepsilon^\alpha = - W
\biggl(\frac{A}{1+W} + \lambda
S^{(1+\alpha)(1+W)}\biggr)^{1/(1+\alpha)} \nonumber \\ &+& A
\biggl(\frac{A}{1+W} + \lambda
S^{(1+\alpha)(1+W)}\biggr)^{-\alpha/(1+\alpha)}. \label{emtmchap1}
\end{eqnarray}
\end{subequations}

The equation for $V$ now reads
\begin{eqnarray}
\ddot V &=& \frac{3\kappa}{2} \Biggl[(1-W)\biggl(\frac{A}{1+W}
V^{(1+\alpha)(1+W)} + \lambda_0 \biggr)^{1/(1+\alpha)} V^{-W}
\nonumber \\ &+& A \biggl(\frac{A}{1+W} V^{(1+\alpha)(1+W)} +
\lambda_0\biggr)^{-\alpha/(1+\alpha)} V^{1 + \alpha (1+W)}\Biggr],
\quad \lambda_0 = \lambda V_0^{(1+\alpha)(1+W)} \label{Vmchap}
\end{eqnarray}
with the solution
\begin{equation}
\int \frac{dV}{\sqrt{C_1 + 3 \kappa V \bigl(\frac{A}{1+W}
V^{(1+\alpha)(1+W)} + \lambda_0 \bigr)^{1/(1+\alpha)} V^{1-W}}} = t
+ t_0, \quad C_1, \, t_0 = {\rm consts}. \label{Vmchapquad}
\end{equation}

Finally, it should be noted that a quintessence with a modified
equation of state
\begin{equation}
p =  W (\ve - \ve_{\rm cr}), \quad W \in (-1,\,0), \label{mq}
\end{equation}
where $\ve_{\rm cr}$ some critical energy density, the spinor field
nonlinearity takes the form
\begin{equation}
F = \lambda S^{1+W} + \frac{W}{1+W}\ve_{\rm cr}. \label{Fmq}
\end{equation}
The spinor field Lagrangian in this case reads
\begin{equation}
L = \frac{i}{2} \biggl[\bp \gamma^{\mu} \nabla_{\mu} \psi-
\nabla_{\mu} \bar \psi \gamma^{\mu} \psi \biggr] - \lambda
S^{(1+W)/2} -\frac{W}{1+W}\ve_{\rm cr}. \label{lspin3}
\end{equation}
Setting $\ve_{\rm cr} = 0$ one gets \eqref{lspin1}. The purpose of
introducing the modified EoS was to avoid the problem of eternal
acceleration. Taking into account that
\begin{subequations}
\begin{eqnarray}
T_0^0 &=& \lambda S^{(1+W)} + \frac{W}{1+W}\ve_{\rm cr}, \label{edmq}\\
T_1^1 = T_2^2 = T_3^3 &=& - \lambda  W S^{(1+W)} +
\frac{W}{1+W}\ve_{\rm cr}, \label{prmq}
\end{eqnarray}
\end{subequations}
for $V$ in this case we find
\begin{equation}
\ddot V = \frac{3\kappa}{2} \Bigl[\lambda V_0^{1 + W} (1 - W) V^{-W}
+ 2W \ve_{\rm cr}V/(1 + W)\Bigr], \label{Vmodq}
\end{equation}
with the solution in quadrature
\begin{equation}
\int \frac{dV}{\sqrt{3 \kappa \bigl[\lambda V_0^{1-W} V^{1 - W} +
W\ve_{\rm cr}V^2/(1 + W)\bigr]  + C_1}} = t + t_0. \label{qdmq}
\end{equation}
Here $C_1$ and $t_0$ are the integration constants. Comparing
\eqref{qdmq} with those with a negative $\Lambda$-term we see that
$\ve_{\rm cr}$ plays the role of a negative cosmological constant.

\subsubsection{Problem of isotropization}

Since the present-day Universe is surprisingly isotropic, it is
important to see whether our anisotropic BI model evolves into an
isotropic FRW model. Isotropization means that at large physical
times $t$, when the volume factor $V$ tends to infinity, the three
scale factors $a_i(t)$ grow at the same rate. Two wide-spread
definition of isotropization read
\begin{subequations}
\label{aniso}
\begin{eqnarray}
{\cal A} &=&  \frac{1}{3} \sum\limits_{i=1}^{3} \frac{H_i^2}{H^2} - 1  \to 0,\\
\Sigma^2 &=& \frac{1}{2}  {\cal A} H^2 \to 0.
\end{eqnarray}
\end{subequations}
Here ${\cal A}$ and $\Sigma^2$ are the average anisotropy and shear,
respectively. $H_i = \dot{a_i}/a_i$ is the directional Hubble
parameter and $H = \dot{a}/a$ average Hubble parameter, where  $a(t)
= V^{1/3}$ is the average scale factor. Here we exploit the
isotropization condition proposed  in \cite{Bronnikov}
\begin{equation}
\frac{a_i}{a}\Bigl|_{t \to \infty} \to {\rm const.} \label{isocon}
\end{equation}
Then by rescaling some of the coordinates, we can make $a_i/a \to
1$, and the metric will become manifestly isotropic at large $t$.

From \eqref{metricf} we find
\begin{eqnarray}
\frac{a_i}{a} = \frac{a_i}{V^{1/3}} =  D_i \exp{\Bigl(X_i \int
\frac{dt}{V}\Bigr)}.\label{isocon1}
\end{eqnarray}
As is seen from \eqref{metricf} in our case $a_i / a \to D_i =$
const as $V \to \infty$. Recall that the isotropic FRW model has
same scale factor in all three directions, i.e., $a_1(t) = a_2(t) =
a_3(t) = a(t)$. So for the BI universe to evolve into a FRW one the
constants $D_i$'s are likely to be identical, i.e., $D_1 = D_2 = D_3
= 1$. Moreover, the isotropic nature of the present Universe leads
to the fact that the three other constants $X_i$ should be close to
zero as well, i.e., $|X_i| << 1$, ($i = 1,2,3$), so that $X_i \int
[V (t)]^{-1}dt \to 0$ for $t < \infty$ (for $V (t) = t^n$ with $n
> 1$ the integral tends to zero as $t \to \infty$ for any $X_i$). It
can be concluded that the spinor field Lagrangian with $W < 1$ leads
to the isotropization of the Universe as $t \to \infty$, moreover,
in case of $W < 0$ the system undergoes an earlier isotropization.

Unfortunately, it is not the end of the story. Due to the specific
behavior of the spinor field in curve space-time there are still
some unresolved questions regarding this case. So before dealing
with other cases let us review this case once again.

In doing so we recall that the expression \eqref{spinor123} can be
rewritten in the form

\begin{equation}
\bp \bg^5 \bg^1 \psi = \bp \bg^5 \bg^2 \psi = \bp \bg^5\bg^3 \psi =
0. \label{spinor1230}
\end{equation}

Recalling that there are 16 independent bilinear combinations:

\begin{subequations}
\label{bf}
\begin{eqnarray}
 S&=& \bar \psi \psi\qquad ({\rm scalar}),   \\
  P&=& \imath \bar \psi \gamma^5 \psi\qquad ({\rm pseudoscalar}), \\
 v^\mu &=& (\bar \psi \gamma^\mu \psi) \qquad ({\rm vector}),\\
 A^\mu &=&(\bar \psi \gamma^5 \gamma^\mu \psi)\qquad ({\rm pseudovector}), \\
T^{\mu\nu} &=&(\bar \psi \sigma^{\mu\nu} \psi)\qquad ({\rm
antisymmetric\,\,\, tensor}),
\end{eqnarray}
\end{subequations}
where $\sigma^{\mu\nu}\,=\,(\imath/2)[\gamma^\mu\gamma^\nu\,-\,
\gamma^\nu\gamma^\mu]$ and 5 invariants, corresponding to these
bilinear forms:
\begin{subequations}
\label{invariants}
\begin{eqnarray}
I &=& S^2, \\
J &=& P^2, \\
I_v &=& v_\mu\,v^\mu\,=\,(\bar \psi \gamma^\mu \psi)\,g_{\mu\nu}
(\bar \psi \gamma^\nu \psi),\\
I_A &=& A_\mu\,A^\mu\,=\,(\bar \psi \gamma^5 \gamma^\mu
\psi)\,g_{\mu\nu}
(\bar \psi \gamma^5 \gamma^\nu \psi), \\
I_T &=& T_{\mu\nu}\,T^{\mu\nu}\,=\,(\bar \psi \sigma^{\mu\nu}
\psi)\, g_{\mu\alpha}g_{\nu\beta}(\bar \psi \sigma^{\alpha\beta}
\psi).
\end{eqnarray}
\end{subequations}
on account of of \eqref{spinor1230} we find $A^1 = A^2 = A^3 = 0$.
Then from the equality
\begin{equation}
A_\mu v^\mu = 0, \label{Av}
\end{equation}
we find
\begin{equation}
A_0 v^0 = \bp  \gamma^5 \gamma_0 \psi \bp \gamma^0 \psi = \bp
\gamma^5 \gamma^0 \psi \bp^* \psi = 0. \label{Av0}
\end{equation}
Since $ \bp^* \psi \ne 0$, from \eqref{Av0} follows that $A^0 = 0$,
hence $I_A = 0$. But according to the Fierz identity  $I_v = - I_A =
I + J$ and $I_T = I - J.$ Hence we obtain
\begin{equation}
I_A = - (S^2 + P^2) =  0, \label{AvSP}
\end{equation}
which leads to the fact that
\begin{equation}
S = \bp \psi = 0, \quad P = i \bp \gamma^5 \psi = 0. \label{SP0}
\end{equation}
This very fact, even without reference to Heisenberg, suggests that
the spinor in this case should be massless.

But the question is whether with $S = 0$ the nonlinearity altogether
vanishes? In case the nonlinearity becomes trivial, we get vacuum
solution, with
\begin{equation}
V = V_1 t + V_2, \quad V_1, V_2 - {\rm consts.} \label{V-0}
\end{equation}
and
\begin{equation}
a_i = D_i \bigl(V_1 t + V_2\bigr)^{\frac{1}{3} + \frac{X_i}{V_1}},
\label{ai-0}
\end{equation}
In this case $\frac{a_i}{a}\Bigl|_{t \to \infty} = \Bigl(V_1 t +
V_2\Bigr)^{X_i/V_1}\Bigl|_{t \to \infty} \nrightarrow {\rm const.}$
It means in absence of nonlinearity no isotropization takes place.

Nevertheless, the case discussed above is worth studying. It shows
how sensitive the spinor field may be to the gravitational one. Now
the question is how to resolve this puzzle?

The problem we are facing now occurs as a result of non-triviality
of the non-diagonal components of the energy momentum tensor of the
spinor field, which imposes some severe restrictions either on
spinor or on gravitational fields. Moreover, this non-triviality is
wholly depends on the affine spinor connections, which is defined by
the gravitational field. One of the possible solutions is to
consider other type of metrics. As it will be shown later, even
imposing total (which corresponds to FRW metric) or partial
(together with spinor field that gives rise to LRS BI metric)
restrictions on the metric functions one can obtain satisfactory
solutions to the problem in question.

But the question is "Is there any way to solve the system within the
scope of BI metric given by \eqref{bi}?" In our view there may be
the following possibilities:

\begin{itemize}

\item Study other types of nonlinearities;\\

\item  Consider spinor fields with larger number of components, such as
dark spinor or ELKO with 8 components and spinors with 16
components;\\

\item  Introduce torsion into the system;

\item  Investigate models with spinor field equations of higher order.

\end{itemize}

It should be noted that in a recent paper  \cite{BIreex} we have
considered the case with with the nonlinear term being some
arbitrary functions  of invariants \eqref{invariants} generated from
bilinear spinor forms \eqref{bf}. Even in that kind of
generalization leads to the conclusion obtained in \eqref{SP0}. So
the nonlinearity should be more general.

Though less likely, there may still be some other way to interpret
the results obtained here. Here is some very close situation, but as
I have already mentioned its probability is very small.
Nevertheless, let us write a few lines about that. It is well known
that the linear spinor field the Lagrangian

\begin{equation}
L_{\rm spl} = \frac{\imath}{2} \biggl[\bp \gamma^{\mu} \nabla_{\mu}
\psi- \nabla_{\mu} \bar \psi \gamma^{\mu} \psi \biggr] - m_{\rm sp}
\bp \psi \label{lspinl}
\end{equation}
vanishes thank to the spinor field equations
\begin{eqnarray}
\imath\gamma^\mu \nabla_\mu \psi - m_{\rm sp} \psi =0, \quad \imath
\nabla_\mu \bp \gamma^\mu +  m_{\rm sp} \bp  = 0. \label{speqln}
\end{eqnarray}
But it does not mean that the Lagrangian is trivial. Or in
Hamiltonian formalism we set, Hamiltonian $H = 0$ that gives
additional constrains, though the Hamiltonian as a whole is
non-trivial. So there might be some extraordinary interpretation of
the nonlinear term being non-trivial, though its arguments becomes
trivial under some specific conditions.

But all these proposals need further detailed investigations. We
plan to study them in some of our forthcoming papers.

\subsection{restrictions on metric functions}

Here we study the other possibility is to keep the components of the
spinor field unaltered. In this case from \eqref{AC} one finds
\begin{equation}
\frac{\dot a_1}{a_1} - \frac{\dot a_2}{a_2} =  \frac{\dot a_2}{a_2}
- \frac{\dot a_3}{a_3} = \frac{\dot a_3}{a_3} - \frac{\dot a_1}{a_1}
= 0, \label{dota1230}
\end{equation}
which can be rewritten as
\begin{equation}
\frac{\dot a_1}{a_1} = \frac{\dot a_2}{a_2} = \frac{\dot a_3}{a_3}
\equiv \frac{\dot a}{a}. \label{dota123}
\end{equation}
Taking into account that
$$\frac{\ddot a_i}{a_i} = \frac{d}{dt}\Bigl(\frac{\dot
a_i}{a_i}\Bigr) + \Bigl(\frac{\dot a_i}{a_i}\Bigr)^2 =
\frac{d}{dt}\Bigl(\frac{\dot a}{a}\Bigr) + \Bigl(\frac{\dot
a}{a}\Bigr)^2 = \frac{\ddot a}{a},$$

the system \eqref{BIE} can be written as a system of two equations:
\begin{subequations}
\label{BIEfr}
\begin{eqnarray}
2\frac{\ddot a}{a} + \frac{\dot
a^2}{a^2}&=&  \kappa T_1^1,\label{11binf}\\
3\frac{\dot a^2}{a^2} &=& \kappa  T_0^0. \label{00binf}
\end{eqnarray}
\end{subequations}

In order to find the solution that satisfies both \eqref{11binf} and
\eqref{00binf} we rewrite \eqref{11binf} in view of \eqref{00binf}
in the following form:
\begin{equation}
\ddot a = \frac{\kappa}{6}\Bigl(3 T_1^1 - T_0^0\Bigr) a. \label{dda}
\end{equation}
Thus in account of non-diagonal components of the spinor field, we
though begin with Bianchi type-I space time, in reality solving the
Einstein field equations for FRW model. Before solving the equation
\eqref{dda}, let us go back to \eqref{metricf}. Taking into account
that
\begin{equation}
\frac{\dot a_i}{a_i} = \frac{\dot V}{3V} + \frac{X_i}{V},
\label{dotai}
\end{equation}
in view of \eqref{dota123} we find that
\begin{equation}
X_1 = X_2 = X_3 = 0. \label{Xi}
\end{equation}
The triviality of the integration constant $X_i$ follows from the
fact that $X_1 + X_2 + X_3 = 0$. Thus the solution \eqref{metricf}
should be written as
\begin{eqnarray}
a_i = D_i V^{1/3} = D_i a, \quad \prod_{i=1}^{3} D_i = 1,
\label{metricf1}
\end{eqnarray}
which means it represents a tiny sector of the general solutions
\eqref{metricf} which one obtains for the BI model in case of
isotropic distribution of matter with trivial non-diagonal
components of energy-momentum tensor, e.g., when the Universe is
filled with perfect fluid, dark energy etc.

Let us now define $a = V^{1/3}$ for different cases. In doing so we
recall that $K$ in this case takes the form
\begin{equation}
S = \frac{a_0^3}{a^3}, \quad a_0 = {\rm const.}. \label{Ka}
\end{equation}

Then then equation for $a$ in case of the spinor field given by
\eqref{lspin1} takes the form

\begin{equation}
\ddot a = - \frac{\kappa \lambda}{6} (1 + 3W) a_0^{3(1+W)}a^{-(2 + 3
W)}, \label{aquint}
\end{equation}
with the solution is quadrature
\begin{equation}
\int \frac{d a}{\sqrt{(\kappa \lambda/3) a_0^{3(1+W)}a^{-(1 + 3 W)}
+ E_1}} = t, \label{qudquint}
\end{equation}
with $E_1$ being integration constant.

\begin{figure}[ht]
\centering
\includegraphics[height=70mm]{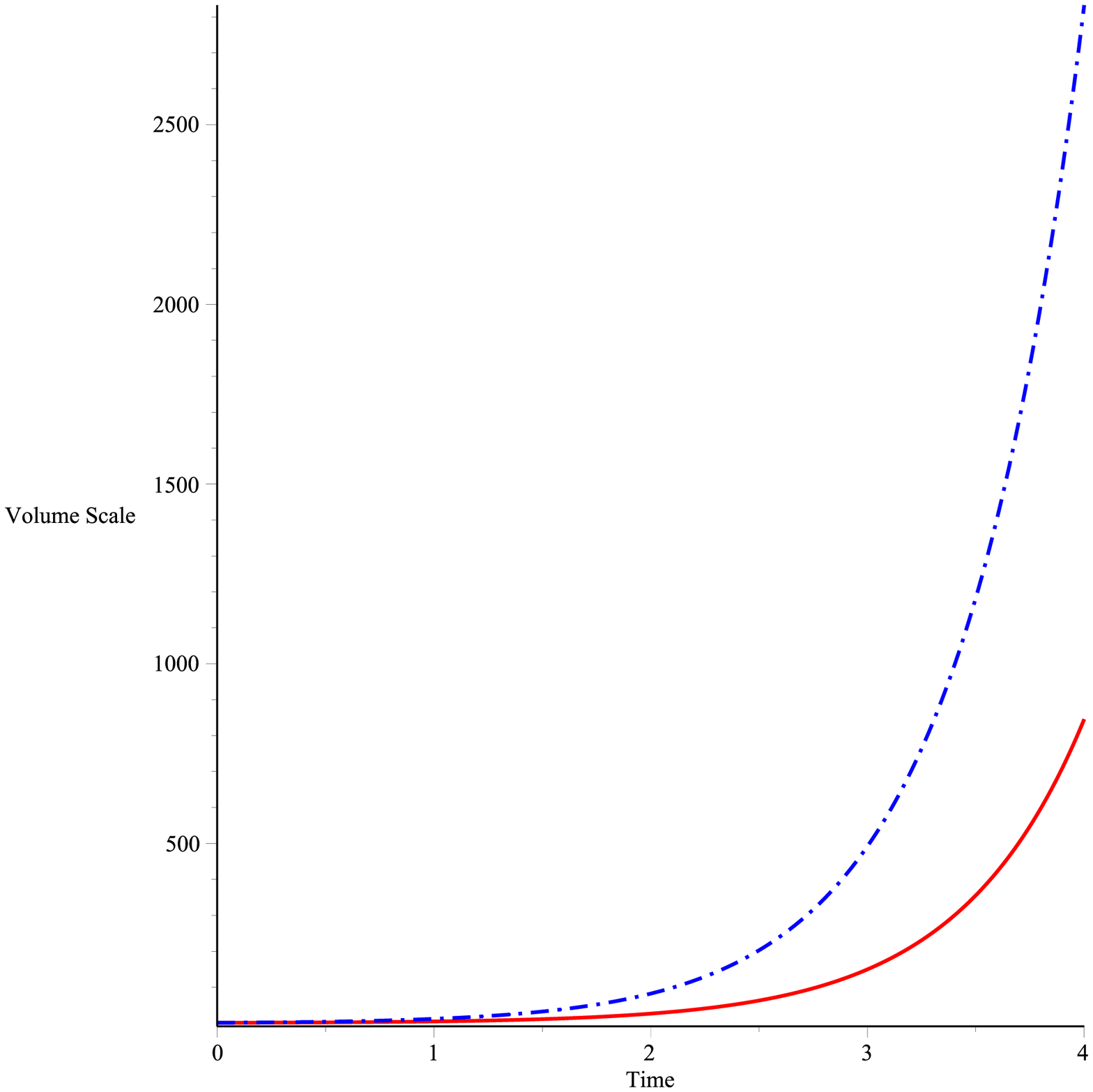} \\
\caption{Evolution of the Universe filled with Chaplygin gas. The
solid (red) line stands for volume scale $V$, while the dash-dot
(blue) line stands for $a^3$. } \label{chap2}.
\end{figure}

As far as Chaplygin scenario is concerned in this case we have

\begin{equation}
\ddot a = - \frac{\kappa}{6} \frac{2A a^{3(1+\alpha)}-
\lambda_0}{a^2 \bigl(A a^{3(1+\alpha)} + \lambda_0
\bigr)^{\alpha/(1+\alpha)}}, \quad \lambda_0 = \lambda
a_0^{3(1+\alpha)} \label{achap}
\end{equation}
This equation is solved numerically and the result is presented in
Fig. \ref{chap2}

\begin{figure}[ht]
\centering
\includegraphics[height=70mm]{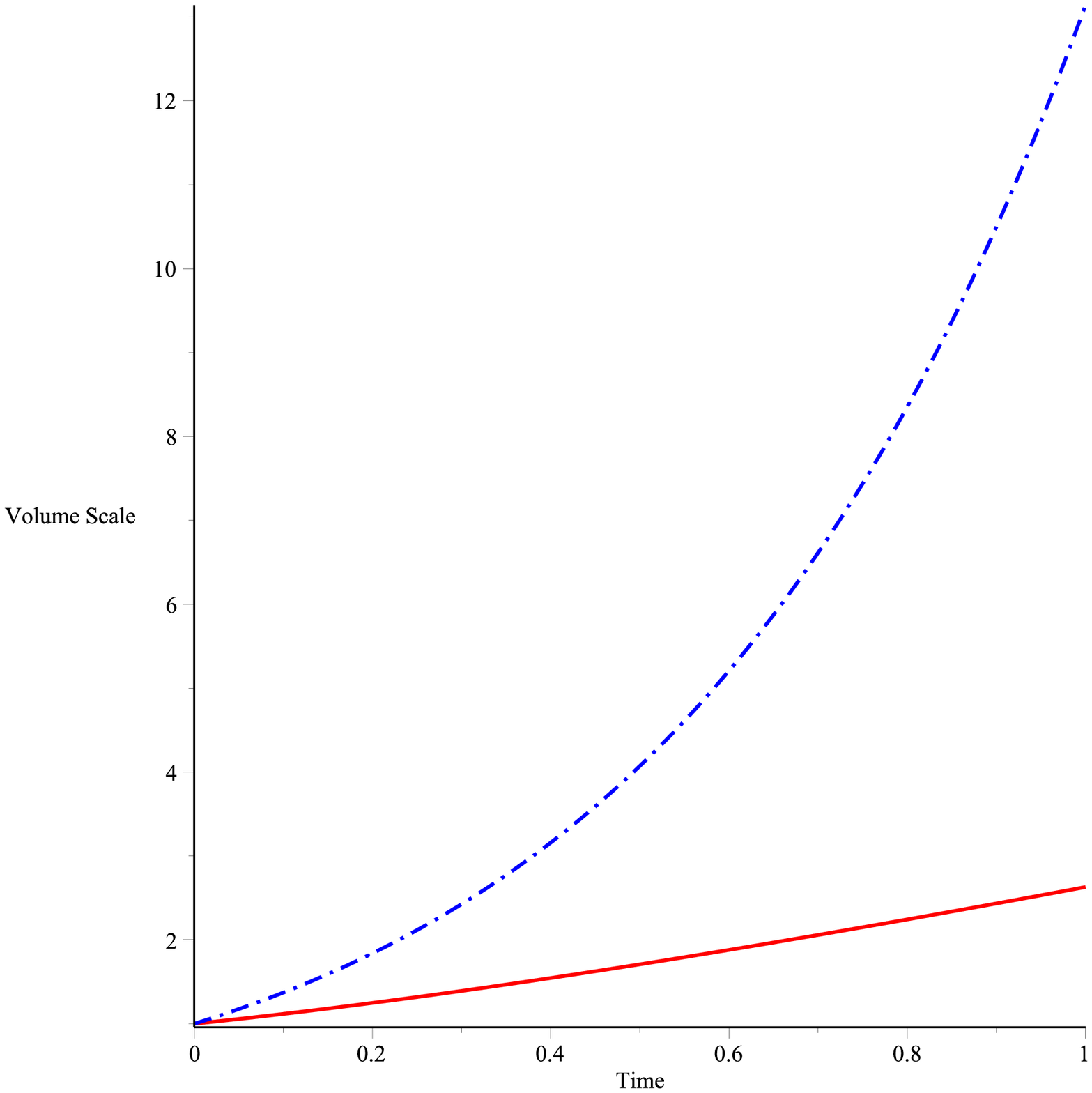} \\
\caption{Evolution of the Universe filled with modified Chaplygin
gas. The solid (red) line stands for volume scale $V$, while the
dash-dot (blue) line stands for $a^3$.} \label{ModChap}.
\end{figure}

For the modified Chaplygin gas we have the following equation for
$a$

\begin{eqnarray}
\ddot a &=& \frac{\kappa}{6} \Biggl[(1-3W)\biggl(\frac{A}{1+W}
a^{3(1+\alpha)(1+W)} + \lambda_0 \biggr)^{1/(1+\alpha)} a^{-(2 +
3W)} \nonumber \\ &+& A \biggl(\frac{A}{1+W} a^{3(1+\alpha)(1+W)} +
\lambda_0\biggr)^{-\alpha/(1+\alpha)} a^{1 + 3\alpha (1+W)}\Biggr],
\quad \lambda_0 = \lambda a_0^{3(1+\alpha)(1+W)} \label{VmchapFRW}
\end{eqnarray}
This equation is solved numerically and the result is presented in
Fig. \ref{ModChap}

Finally we consider the case with modified quintessence. Inserting
\eqref{edmq} and \eqref{prmq} into \eqref{dda} in this case we find
\begin{equation}
\ddot a  = -\frac{\kappa}{6}\Bigr[(3W+1)\lambda a_0^{3(1+W)}
a^{-(3W+2)} - \frac{2W}{1+W}\ve_{\rm cr} a\Bigr], \label{FRWmq}
\end{equation}
with he solution
\begin{equation}
\int\frac{d a}{\sqrt{(\kappa/3)\Bigl[\lambda a_0^{3(1+W)} a^{-(3W
+1)} + [W/(1+W)] \ve_{\rm cr} a^2 + E_2\Bigr]}} = t, \quad E_2 =
{\rm const}. \label{dda2}
\end{equation}
It can be shown that in case of modified quintessence the pressure
is sign alternating. As a result we have a cyclic mode of evolution.

It should be noted that the metric functions $a_i$ in this case not
necessarily be identic, rather one can write
\begin{equation}
a_1 = c_1 a, \quad a_2 = c_2 a, \quad a_3 = c_3 a, \quad c_1 c_2 c_3
= 1, \label{ais}
\end{equation}
with $c_i$ being some integration constants.

As far as isotropization is concerned, in this case from \eqref{ais}
we find
\begin{eqnarray}
\frac{a_i}{a} = c_i = {\rm const.}, \label{isoconfr}
\end{eqnarray}
for any given time. Though the solutions for metric functions can be
obtained solving Einstein equations for FRW space-time, depending on
the constants it might not be isotropic from the very beginning, but
in the course of time becomes isotropic. For the metric to be
completely isotropic the constants $c_i$ should be identical, i.e.,
$c_1 = c_2 = c_3$.

\begin{figure}[ht]
\centering
\includegraphics[height=70mm]{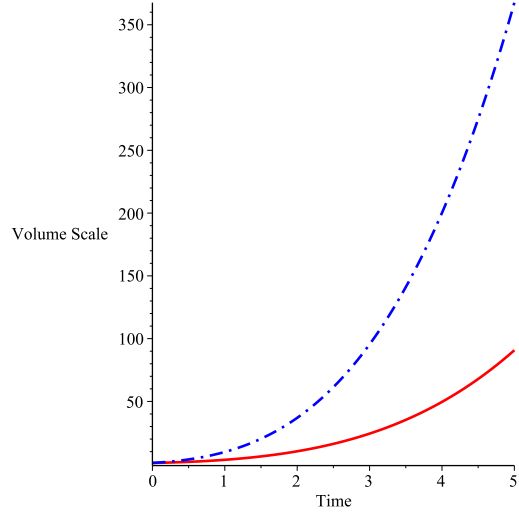} \\
\caption{Evolution of the Universe filled with quintessence. The
solid (red) line stands for volume scale $V$, while the dash-dot
(blue) line stands for $a^3$.} \label{quint2}.
\end{figure}

\begin{figure}[ht]
\centering
\includegraphics[height=70mm]{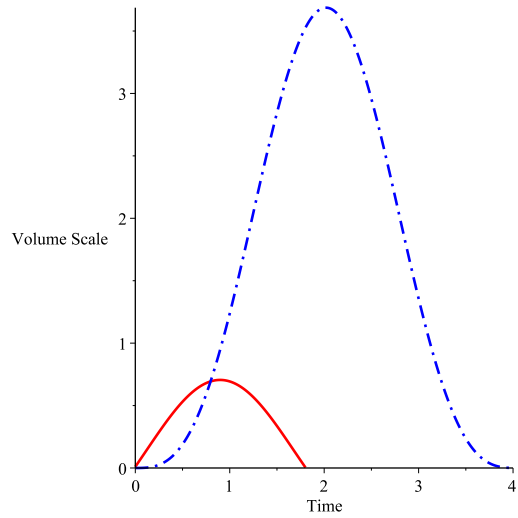} \\
\caption{Evolution of the Universe filled with quintessence with
modified equation of state. The solid (red) line stands for volume
scale $V$, while the dash-dot (blue) line stands for $a^3$. }
\label{mq2}.
\end{figure}

In what follows we illustrate the evolution of the Universe filled
with quintessence, Chaplygin gas and quintessence with modified
equation of state for two different cases: when the restrictions are
imposed on the metric functions and when both spinor field and the
metric functions were restricted. In Figures \ref{quint2},
\ref{chap2}, \ref{ModChap} and \ref{mq2} we illustrated the
evolution of the Universe filled with quintessence, Chaplygin gas,
modified Chaplygin gas and quintessence with modified equation of
state, respectively. The solid (red) line stands for the volume
scale, when the restrictions were imposed on both the components of
the spinor field and the metric functions. In this case the
spacetime is given by a LRS BI model and the isotropization takes
place asymptotically. The blue line shows the evolution of the
Universe when the restrictions were impose on the metric functions.
In this case the spacetime becomes isotropic from the very beginning
and is described by a FRW cosmological model. Here we plot the
volume scale as $a^3$, which $a$ being the average scale factor.

As one sees, in case of early isotropization the Universe grows
rapidly.

\subsection{restrictions on metric functions and Spinor field}

As we have already seen, the restrictions imposed only one the
spinor field may lead to vacuum solution, while the restrictions of
metric functions give rise to isotropic FRW model from the very
beginning.

As a third path one may offer a model by imposing restrictions on
both metric functions and spinor field. In doing so, let us assume,
say
\begin{equation}
\frac{\dot a_2}{a_2} - \frac{\dot a_3}{a_3} = 0, \label{A23}
\end{equation}
In this case we get a LRS Bianchi type-I space-time with the metric
\begin{equation}
ds^2 = dt^2 - a_1^{2} \,dx^{2} - a_2^{2} \bigl[dy^2 + \,dz^2\bigr].
\label{lrsbi}
\end{equation}
Restrictions on the spinor field in this case look
\begin{equation}
\bp \bg^5 \bg^2 \psi = \bp \bg^25\bg^3 \psi = 0. \label{spinor523}
\end{equation}
Such type of restriction was earlier used in \cite{FabJMP}. Under
this condition we now have the following relations between the
coefficients of spinor field
\begin{subequations}
\label{spconsnew1234}
\begin{eqnarray}
C_1^* C_2 - C_3^* C_4 &=&  C_2^* C_1 - C_4^* C_3,
\label{spconsnew12}\\
C_1^* C_1 - C_3^* C_3 &=&  C_2^* C_2  - C_4^* C_4 = \frac{V_0}{2}.
\label{spconsnew34}
\end{eqnarray}
\end{subequations}

The Einstein Equations in this case read
\begin{subequations}
\label{BIELRS}
\begin{eqnarray}
2\frac{\ddot a_2}{a_2} + \Bigl(\frac{\dot
a_2}{a_2}\Bigr)^2 &=&  \kappa \bigl( F(S) - S F_S\bigr),\label{11lrsbi}\\
\frac{\ddot a_1}{a_1} +\frac{\ddot a_2}{a_2} + \frac{\dot
a_1}{a_1}\frac{\dot a_2}{a_2}&=&  \kappa \bigl( F(S) - S F_S\bigr),\label{33lrsbi}\\
2\frac{\dot a_1}{a_1}\frac{\dot a_2}{a_2} + \Bigl(\frac{\dot
a_2}{a_2}\Bigr)^2&=& \kappa F(S), \label{00lrsbi}
\end{eqnarray}
\end{subequations}
where we set the spinor mass $m_{\rm sp} = 0$. As in previous case,
defining
\begin{equation}
V = a_1 a_2^2, \label{Vlrs}
\end{equation}
we find
\begin{equation}
a_1 = D^2 V^{1/3} \exp[2X \int\frac{dt}{V}], \quad a_2 = (1/D)
V^{1/3} \exp[-X \int\frac{dt}{V}], \label{a1a2}
\end{equation}
with $D$ and $X$ being some arbitrary constants of integration. As
far as $V$ is concerned, summation of \eqref{11lrsbi}, 2 times
\eqref{33lrsbi} and 3 times \eqref{00lrsbi} gives
\begin{equation}
\ddot V = \frac{3 \kappa}{2} \bigl(2F(S) - S F_S\bigr)\,V.
\label{Vlrsbi}
\end{equation}
Further choosing the nonlinear term in the forms \eqref{sol1},
\eqref{chapsp}, and \eqref{Fmq}, respectively, we obtain the
analogical solutions as in first case. And as in that case here too
we found that the isotropization process takes place asymptotically.

\section{Physical aspects of the models}

In this section we discuss the physical aspects of the models
considered above. Since the Bianchi type - I model within the spinor
source needs further considerations, we begin with the FRW like
case, when restrictions were imposed on the metric functions only.
Let us note that though perfect fluid, quintessence etc. given by
\eqref{zeta}, \eqref{chap}, \eqref{mchap} and \eqref{mq} can be
simulated by the spinor field, it does not necessarily mean that we
should confine to those models only. In this section we study the
spinor field with a non-zero massive term and compare the results
obtained with some recent observations. To begin with we write the
EoS parameter $\omega$ as a ration between the pressure and energy
density. Taking into account that $T_0^0 = \ve$ and $T_1^1 = -p$
from \eqref{BIEfr} one obtains

\begin{equation}
\omega = \frac{p}{\ve}  = - \frac{2 \frac{\ddot a}{a} + \frac{\dot
a^2}{a^2}}{3 \frac{\dot a^2}{a^2}}, \label{omega01}
\end{equation}
which on account of $q = - a{\ddot a}/{\dot a}^2$ gives the well
known relation between the deceleration parameter $q$ and EoS
parameter $\omega$:
\begin{equation}
q = \frac{3}{2}\bigl(\omega + \frac{1}{3}\bigr). \label{omegaq}
\end{equation}
From \eqref{omegaq} we see, for $\omega > - 1/3$  the Universe
expands with deceleration, while an accelerative mode of expansion
takes place only when $\omega < - 1/3$. For $\omega = - 1/3$ the
deceleration parameter becomes trivial. In this case the metric
function is either constant $a = const.$ or a linear function of
time $a = C_1 t + C_2$.

On the other hand, from \eqref{Ttot00} and \eqref{Ttot11} one finds

\begin{equation}
\omega = \frac{p}{\ve}  = \frac{S F_S - F}{mS + F}.\label{omega02}
\end{equation}

In our previous papers \cite{saha2001a,saha2004a} we considered
different types of spinor field nonlinearities. But given the fact
that most of the established source fields are simulated by the
spinor field nonlinearities given as power law, here we consider
only that case setting  $F = \lambda S^n$. In this case we find
\begin{equation}
\omega = \frac{\lambda (n - 1) S^n}{mS + \lambda
S^n}.\label{omega03}
\end{equation}
The relations \eqref{omega03} shows that only for $ n < 2/3$ the
spinor field in the given model can give rise to an accelerated mode
of expansion.

Further recalling that $S = a_0^3 / a^3$ from \eqref{omega03} we
find
\begin{equation}
\omega = \frac{\lambda (n - 1)}{\lambda  + {\tilde m} a^{3(n - 1)}},
\quad {\tilde m} = m/a_0^{3(n - 1)}.\label{omega04}
\end{equation}
From \eqref{omega04} we see that for a massless spinor field the EoS
parameter is a constant, namely $\omega = n - 1$, while if the
spinor field has a nontrivial mass, the EoS parameter is
time-dependent. Moreover, in absence of nonlinear term ($\lambda =
0$ and/or $n = 1$) the EoS parameter becomes trivial ($\omega = 0$).
Note that this conclusion in no way contradicts our previous
results. In order to simulate different matters the EoS parameter in
\eqref{beos} was taken to be constant. This very assumption leads to
the spinor field Lagrangian with $m_{\rm sp} = 0$.

In what follows we study the case when the EoS parameter is time
dependent. It can be a function of red-shift $z$ or scale factor $a$
(which it indeed is) as well. The red-shift dependence of $\omega$
can be linear like
\begin{equation}
 \omega(z) =  \omega_{0} + \omega^{'} z,\label{redsh}
\end{equation}
 with $\omega^{'}$ = $\frac{d\omega}{dz}|_{z=0}$ (see Refs. \cite{Huterer,Weller}
 or nonlinear as \cite{Chevallier,Linder}
\begin{equation}
\omega(z) = \omega_{0} + \frac{\omega_{1}z}{1 + z}. \label{redshnl}
\end{equation}
 So, as for as the scale
factor dependence of $\omega$ is concern, the parametrization
\begin{equation}
\omega(a) = \omega_{0} + \omega_{a}(1 - a), \label{redshpar}
\end{equation}
where $\omega_{0}$ is the present value ($a = 1$) and $\omega_{a}$
is the measure of the time variation $\omega^{'}$ is widely used in
the literature \cite{Linder1}.

So, if the present work is compared with experimental results
obtained in \cite{Knop,Tegmark1,Hinshaw,Komatsu}, then one can
conclude that the limit of $\omega$ provided by equation
\eqref{omega04} may accommodated with the acceptable range of EoS
parameter. As it was already noticed, the EoS parameter vanishes in
absence of spinor field nonlinearity.

For the value of $\omega$ to be in consistent with observation
\cite{Knop}, we have the following general condition
\begin{equation}
 a_{[1]} < a < a_{[2]}, \label{a12}
\end{equation}
where
\begin{equation}
 a_{[1]} = \biggl[-\frac{(n + 0.67)\lambda}{1.67 {\tilde
 m}}\biggr]^{1/3(n-1)}, \quad   a_{[2]} = \biggl[-\frac{(n - 0.38)\lambda}{0.62 {\tilde
 m}}\biggr]^{1/3(n-1)} \label{knopFRW}
\end{equation}

For this constrain, we obtain  $-1.67 < \omega < -0.62$, which is in
good agreement with the limit obtained from observational results
coming from SNe Ia data \cite{Knop}.

For the value of $\omega$ to be consistent with observation
\cite{Tegmark1}, we have the following general condition
\begin{equation}
a_{[3]} < a < a_{[4]}, \label{a34}
\end{equation}
where

\begin{equation}
 a_{[3]} = \biggl[-\frac{(n + 0.33)\lambda}{1.33 {\tilde
 m}}\biggr]^{1/3(n-1)}, \quad  a_{[4]} = \biggl[-\frac{(n - 0.21)\lambda}{0.79 {\tilde
 m}}\biggr]^{1/3(n-1)}. \label{Tegmark1FRW}
\end{equation}

For this constrain, we obtain  $-1.33 < \omega < -0.79$, which is in
good agreement with the limit obtained from observational results
coming from SNe Ia data \cite{Tegmark1}.

For the value of $\omega$ to be consistent with observation
\cite{Hinshaw,Komatsu}, we have the following general condition
\begin{equation}
a_{[5]} < a < a_{[6]},\label{a56}
\end{equation}
where
\begin{equation}
 a_{[5]} = \biggl[-\frac{(n + 0.44)\lambda}{1.44 {\tilde
 m}}\biggr]^{1/3(n-1)}, \quad  a_{[6]} = \biggl[-\frac{(n - 0.08)\lambda}{0.92 {\tilde
 m}}\biggr]^{1/3(n-1)}. \label{HinshawFRW}
\end{equation}

For this constrain, we obtain  $-1.44 < \omega < -0.92$, which is in
good agreement with the limit obtained from observational results
coming from SNe Ia data \cite{Hinshaw,Komatsu}.

We also observed that if
\begin{equation}
 a_{[0]} = \biggl[-\frac{n\lambda}{{\tilde
 m}}\biggr]^{1/3(n-1)}, \label{Caldwell}
\end{equation}
then for $a = a_{[0]}$ we have $\omega = -1$, i.e., we have universe
with cosmological constant. If $a < a_{[0]}$ the we have $\omega >
-1$ that corresponds to quintessence, while for $a > a_{[0]}$ we
have $\omega < -1$, i.e., Universe with phantom matter
\cite{Caldwell1}.

Since for the Bianchi type model given by \eqref{bi} both the spinor
mass and spinor field nonlinearity vanish, there is no need to carry
out the foregoing analysis for this case. As far as LRS Bianchi
type-I metric is concerned, one can compare the result with
observational data in the same way, as it is done for FRW case. In
this case $S = V_0 / V$ from \eqref{omega03} we find
\begin{equation}
\omega = \frac{\lambda (n - 1)}{\lambda  + {\tilde m} V^{(n - 1)}},
\quad {\tilde m} = m/V_0^{(n - 1)}.\label{omega05}
\end{equation}

For the value of $\omega$ to be in consistent with observation
\cite{Knop}, we have the following general condition
\begin{equation}
 V_{[1]} < V < V_{[2]}, \label{V12}
\end{equation}
where
\begin{equation}
 V_{[1]} = \biggl[-\frac{(n + 0.67)\lambda}{1.67 {\tilde
 m}}\biggr]^{1/(n-1)}, \quad   V_{[2]} = \biggl[-\frac{(n - 0.38)\lambda}{0.62 {\tilde
 m}}\biggr]^{1/(n-1)} \label{knopLRSBI}
\end{equation}

For this constrain, we obtain  $-1.67 < \omega < -0.62$, which is in
good agreement with the limit obtained from observational results
coming from SNe Ia data \cite{Knop}.

For the value of $\omega$ to be consistent with observation
\cite{Tegmark1}, we have the following general condition
\begin{equation}
V_{[3]} < V < V_{[4]}, \label{V34}
\end{equation}
where

\begin{equation}
 V_{[3]} = \biggl[-\frac{(n + 0.33)\lambda}{1.33 {\tilde
 m}}\biggr]^{1/(n-1)}, \quad  V_{[4]} = \biggl[-\frac{(n - 0.21)\lambda}{0.79 {\tilde
 m}}\biggr]^{1/(n-1)}. \label{Tegmark1LRSBI}
\end{equation}

For this constrain, we obtain  $-1.33 < \omega < -0.79$, which is in
good agreement with the limit obtained from observational results
coming from SNe Ia data \cite{Tegmark1}.

For the value of $\omega$ to be consistent with observation
\cite{Hinshaw,Komatsu}, we have the following general condition
\begin{equation}
V_{[5]} < V < V_{[6]},\label{V56}
\end{equation}
where
\begin{equation}
 V_{[5]} = \biggl[-\frac{(n + 0.44)\lambda}{1.44 {\tilde
 m}}\biggr]^{1/(n-1)}, \quad  V_{[6]} = \biggl[-\frac{(n - 0.08)\lambda}{0.92 {\tilde
 m}}\biggr]^{1/(n-1)}. \label{HinshawLRSBI}
\end{equation}

For this constrain, we obtain  $-1.44 < \omega < -0.92$, which is in
good agreement with the limit obtained from observational results
coming from SNe Ia data \cite{Hinshaw,Komatsu}.

We also observed that if
\begin{equation}
 V_{[0]} = \biggl[-\frac{n\lambda}{{\tilde
 m}}\biggr]^{1/(n-1)}, \label{CaldwellLRSBI}
\end{equation}
then for $V = V_{[0]}$ we have $\omega = -1$, i.e., we have universe
with cosmological constant. If $V < V_{[0]}$ the we have $\omega >
-1$ that corresponds to quintessence, while for $V > V_{[0]}$ we
have $\omega < -1$, i.e., Universe with phantom matter
\cite{Caldwell1}.

\section{Conclusion}

Within the scope of Bianchi type-I space time we study the role of
spinor field on the evolution of the Universe. It is shown that even
in case of space independence of the spinor field it still possesses
non-zero non-diagonal components of energy-momentum tensor thanks to
its specific relation with gravitational field. This fact plays
vital role on the evolution of the Universe. There might be three
different scenarios.

In the first case only the components of the spinor field are
affected leaving the space-time initially anisotropic that evolves
into an isotropic one asymptotically. Unfortunately, due to the
specific behavior of the spinor field the bilinear forms constructed
from it becomes trivial, thus giving rise to a massless and linear
spinor field Lagrangian. So this case presents a very tiny sector of
spinor field.

According to the second scenario, where restrictions were imposed
wholly on metric functions, they comes out to be proportional to
each other right from the beginning,i.e.,
\begin{equation}
a_1 \sim a_2 \sim a_3, \label{a123sim}
\end{equation}
and can be completely described by the Einstein field equations for
FRW metric. As numerical analysis shows, in the second case the
Universe expands rather rapidly that leads to the early
isotropization of spacetime.

A third possibility was considered when the non-diagonal components
of energy-momentum tensor influence both the spinor field and metric
functions simultaneously. This case is described by a locally
rotationally symmetric Bianchi type-I (LRS-BI) spacetime. In this
case isotropization takes place asymptotically and the nonlinearity
remains non-trivial.

The results obtained were compared to the recent observational data
and the acceptable ranges for the EoS parameter were established. It
was found that if the relation between the pressure and energy
density obeys a barotropic equation of state, only a non-trivial
spinor mass can give rise to a dynamic EoS parameter.

It should be noted that in case when the restrictions are imposed
only on the components of the spinor field, though the system is
solved completely, the bilinear spinor forms become trivial. So we
need some alternative approach to this problem. Since this problem
occurs due to the non-diagonal components of the energy momentum
tensor of the spinor field which is directly related to spinor
affine connection, it needs a very careful treatment. We plan to
address this problem in some of our coming papers.

\vskip 0.1
cm

\noindent {\bf Acknowledgments}\\
This work is supported in part by a joint Romanian-LIT, JINR, Dubna
Research Project, theme no. 05-6-1119-2014/2016. Taking the
opportunity I would also like to thank the reviewers for some
helpful discussions and references.

\end{document}